\documentclass[12pt,preprint]{aastex}

\shorttitle{Evolving to SNe Ia with long delay time}
\shortauthors{Chen \& Li}

\begin{document}


\title{Evolving to type Ia supernovae with long delay time}


\author{Wen-Cong Chen \altaffilmark{1,2}  and Xiang-Dong Li \altaffilmark{1}}
\altaffiltext{1}{Department of Astronomy, Nanjing University,
    Nanjing 210093, China; chenwc@nju.edu.cn,lixd@nju.edu.cn}
\altaffiltext{2}{Department of Physics, Shangqiu Normal
University, Henan, 476000, China}



\begin{abstract}
Recent investigations on the delay time of type Ia supernovae have
set useful constraints on the progenitors of type Ia supernovae.
Here we have calculated the evolution of close binaries consisting
of a white dwarf and a main-sequence or subgiant companion. We
assume that, once Roche lobe overflow occurs a small fraction of
the lost mass from the system forms a circumbinary disk, which
extracts the orbital angular momentum from the system through
tidal torques. Our calculations indicate that the existence of
circumbinary disk can enhance the mass transfer rate and cause
secular orbital shrinkage. The white dwarf can grow in mass
efficiently to trigger type Ia supernovae even with relatively
low-mass ($\la 2 M_{\odot}$) donor stars. Thus this scenario
suggest a new possible evolutionary channel to those type Ia
supernovae with long delay time $\sim 1-3$ Gyr.
\end{abstract}


\section{Introduction}
Many works have suggested that type Ia supernovae (SNe Ia) can be
used as the standard candlelight to determine the cosmological
distances, providing the strong evidence that the low red-shift
Universe is accelerating \citep[e.g.][]{ries98,perl99,ries04}. It is
widely believed that SNe Ia are thermonuclear explosions of
accreting CO white dwarfs (WDs) when their masses grow beyond a
critical mass \citep{hoy60}. However, the nature of the progenitors
and the related accretion processes have still remained unclear.
Several possible evolutionary scenarios for SN Ia explosions have
been proposed so far. At present, there exist Chandrasekhar (Ch)
mass model \citep{woos86} and sub-Chandrasekhar (sub-Ch) mass model
\citep{nomo82b} for the critical mass of the exploding WDs.
Furthermore, both double degenerate model (DD) \citep{iben84,webb84}
and single degenerate model (SD) \citep{nomo82a} have been also
proposed as the candidates for progenitors of SNe Ia \citep[for a
review see][]{bran95}. For the DD model, previous works indicated
that the expected accretion rates may cause the accretion-induced
collapse (AIC) of the WDs and the formation of neutron stars instead
of SN Ia explosions \citep{nomo85,saio85,timm94}.

Binary evolution investigations adopting the SD/Ch scenarios
obtained two types of progenitors of SNe Ia. One is close binaries
with a main-sequence or subgiant companion of mass $\sim
2-3.5M_{\odot}$ and an initial orbital period of $\sim$ 1 d
\citep{li97,lang00,han04}. The required accretion rate for steady
nuclear burning onto WDs could be obtained with thermal-timescale
mass transfer from the more massive donor stars. The other is wide
binaries with a red giant of masses $\sim1 M_{\odot}$ and an
initial orbital period of $\sim$ 100 d \citep{hach96,li97}.
However, mass accretion in these systems may be unstable due to
the thermal-viscous instability in the accretion disks
\citep{cann93}.

The various progenitor scenarios can be tested by comparison of
the expected and measured distribution of the time delay between
the formation of the progenitor systems and their explosions.
Recent high-$z$ supernova observations suggested the mean delay
time of $t\sim 2-4$ Gyr \citep{stro04,stro05,gal04,barr06,fors06}.
However, the main-sequence lifetime of $\ga2M_{\odot}$ star should
be $\la$ 1 Gyr in most of the SD/DD SNe Ia progenitor scenarios
\citep{han04}, and the mass transfer timescale ($\la 100\rm Myr$)
between the components of binary can be neglected. This led
\citet{bel05} to conclude that only SN Ia progenitors in the DD
scenario has a characteristic delay time of $\sim 3$ Gyr.
Therefore, if the SD/Ch scenario really works for (at least part
of) SNe Ia, a progenitor system with a relatively low-mass donor
star is needed.

Supersoft X-ray sources (SSS), originally discovered by the
\emph{Einstein} satellite in the Large Magellanic Cloud
\citep{long81}, may be the observational evidence of the
progenitors of SNe Ia \citep{kah97}. A steady-state nuclear
burning model on the surface of the accreting WD in a binary was
used to interpreted the SSS CAL 83 and CAL 87 by \citet{heu92},
who suggested that a near-main-sequence secondary star with a mass
of $1.3-2.5 M_{\odot}$ could provide the required accretion rate
for steady nuclear burning via thermal-timescale mass transfer.
However, the catalysmic variable SSS J0439.8$-$6809,
J0537.7$-$7304 and 1E 0035.4$-$7230 have very small mass ratios
and orbital periods ($\sim 3-4$ hr) \citep[][and references
therein]{spru01}, which are significantly less than that required
for a binary with thermal-timescale mass transfer at a rate $\sim
10^{-7}\,M_{\odot}$yr$^{-1}$. \citet{tees98} proposed that there
may exist an efficient mechanism (irradiation-driven mass loss) to
enhance the mass transfer rates in these SSS.

The purpose of this paper is to explore the possible progenitor
systems for SNe Ia with delay times of a few Gyr in the SD/Ch model.
Enlightened by the original works of \citet{spru01} and
\citet{taam01}, we consider the orbital angular momentum loss
through the tidal interaction of binary system with a circumbinary
(CB) disk during the mass transfer in a close binary. Our previous
works indicate that the CB disk is an efficient mechanism extracting
angular momentum from the binary system \citep{chen06a,chen06b},
which may enhance the mass transfer rates and help mass accumulation
on the WD. The prescriptions of binary evolution calculations are
described in section 2. In section 3, we present the numerically
calculated results for the evolutionary sequences of the WD
binaries. Finally, we discuss and summarize our results in section
4.

\section{Model}

We consider a binary system consisting of a CO WD (of mass $M_{\rm
WD}$), and a main-sequence or subgiant companion (of mass $M_{\rm
d}$) with solar chemical composition ($X=0.7$, $Y=0.28$, $Z=0.02$).
Mass transfer will occur via Roche lobe overflow of the companion
due to nuclear expansion of the star or orbital shrinkage.
We have calculated the evolution of WD binaries adopting an updated
version of the stellar evolution code developed by
\citet{eggl71,eggl72} \citep[see also][]{han94,pols95}. In the
calculations we take the ratio of the mixing length to the pressure
scale height to be 2.0. We include the following mass loss processes
and orbital angular momentum loss mechanisms during the mass
exchange.

\subsection{Mass accumulation efficiency}
The key factor for the growth of the WD mass is the accumulation
ratio $\alpha$ of the accreted hydrogen converted into heavier
elements. Unfortunately, there are large uncertainties in estimating
the values of $\alpha$. For the fraction $\alpha_{\rm H}$ of the
transferred mass during hydrogen burning, we adopt the prescription
by \citet{hach99} and \citet{han04}. If the mass transfer rate
$\dot{M_{\rm d}}$ is higher than a critical value $\dot{M_{\rm
cr}}$, we assume that hydrogen is converted into helium at a rate
limited to $\dot{M_{\rm cr}}$ due to the strong optically thick
winds from the WD. The critical mass transfer rate can be written as
\begin{equation}
\dot{M_{\rm cr}}\simeq 5.3\times
10^{-7}\left(\frac{1.7-X}{X}\right)\left(\frac{M_{\rm
WD}}{1M_{\odot}}-0.4\right) M_{\odot}\rm yr^{-1},
\end{equation}
where $X$ is the hydrogen mass abundance of the accreted matter.
Below this value, all hydrogen is assumed to be burned into helium,
until the mass transfer rate becomes less than $\dot{M_{\rm cr}}/8$,
and strong hydrogen shell flashes occur \citep{kove94}. Thus
\begin{equation}
 \alpha_{\rm H}= \left\{ \begin{array}{l@{\quad,\quad}l}
-\dot M_{\rm cr}/\dot{ M_{\rm d}} &  -\dot{M_{\rm d}}>\dot{M}_{\rm cr},\strut\\
1 &  \dot{M}_{\rm cr}\ge -\dot{M_{\rm d}}\ge \dot{M}_{\rm cr}/8,\strut\\
0 &  -\dot{M_{\rm d}}< \dot{M}_{\rm cr}/8.\strut\\
\end{array} \right.
\end{equation}
After the gradual increase of the mass in the helium layer on the
surface of the WD, helium ignition occurs. A part of the envelope
mass is expected to be blown off due to the helium-shell flashes
\citep{kato89}. For the helium mass accumulation ratio $\alpha_{\rm
He}$, we adopt the prescription given by \citet{hach99},
\begin{equation}
\alpha_{\rm He}= \left\{ \begin{array}{l}
-0.175\, (\log \dot M_{\rm He}+5.35)^2+1.05, \\
\begin{array}{l@{\quad\quad\quad\quad\quad\quad}l}
& -7.3 < \log  \dot M_{\rm He} <-5.9, \\
1\quad , &  -5.9 \le \log  \dot M_{\rm He} \le -5.\\
\end{array} \end{array}\right.
\end{equation}
Summarize the above prescriptions, the mass growth rate of the CO
WD is $ \dot{M}_{\rm WD}= -\alpha_{\rm H}\alpha_{\rm He}
\dot{M_{\rm d}}$. The mass lost rate from the binary system can
then be written as $ \dot{M}=(1-\alpha_{\rm H}\alpha_{\rm
He})\dot{M_{\rm d}}$.

\subsection{Orbital angular momentum losses}

During the mass exchange in WD binaries some fraction of the
transferred matter from the donor star may be lost from the system
in various ways. For example, high velocity outflows were observed
from the ultraviolet spectra of RX J0513.9-6951 and CAL 83
\citep{gans98}. Part of the lost matter may form a disk structure
surrounding the binary system rather leave the binary system
\citep{van73,van94}. The optical and UV spectrum of the SSS RX
J0019.8$+$2156 indicate the presence of circumbinary material
\citep{kud02,hut01}. Here we assume that a small fraction $\delta$
of the mass lost feeds into the CB disk surrounding the binary
system at its inner radius $r_{\rm i}$. Tidal torques are then
exerted on the CB disk via gravitational interaction, extracting
orbital angular momentum from the binary system
\citep{spru01,taam01}. The angular momentum loss rate via the CB
disk is \citep{chen06b}
\begin{equation}
\dot{J}_{\rm CB}=\gamma\left(\frac{2\pi a^2}{P_{\rm
orb}}\right)\delta\dot{M}\left(\frac{t}{t_{\rm vi}}\right)^{1/3},
\end{equation}
where $\gamma^{2}=r_{\rm i}/a$, $t$ is the mass transfer time,
$P_{\rm orb}$ and $a$ are the orbital period and the separation of
the binary, respectively. In the standard "$\alpha$ viscosity"
prescription \citep{Shak73}, the viscous timescale $t_{\rm vi}$ at
the inner edge in the CB disk is given by $ t_{\rm
vi}=\frac{2\gamma^{3}P_{\rm orb}}{3\pi\alpha_{\rm SS}\beta^{2}}, $
where $\beta=H_{\rm i}/r_{\rm i}$, $\alpha_{\rm SS}$ and $H_{\rm
i}$ are the viscosity parameter and the scale height of the CB
disk, respectively.

In addition, we assume that the other part $(1-\delta)$ of the lost
mass $\dot{M}$ is ejected in the vicinity of the WD in the form of
isotropic winds or outflows, carrying away the specific orbital
angular momentum of the WD. The mass loss in the donor's wind and
its effect on the change of orbital angular momentum are
comparatively negligible.

\begin{figure}
\epsscale{.95} \plotone{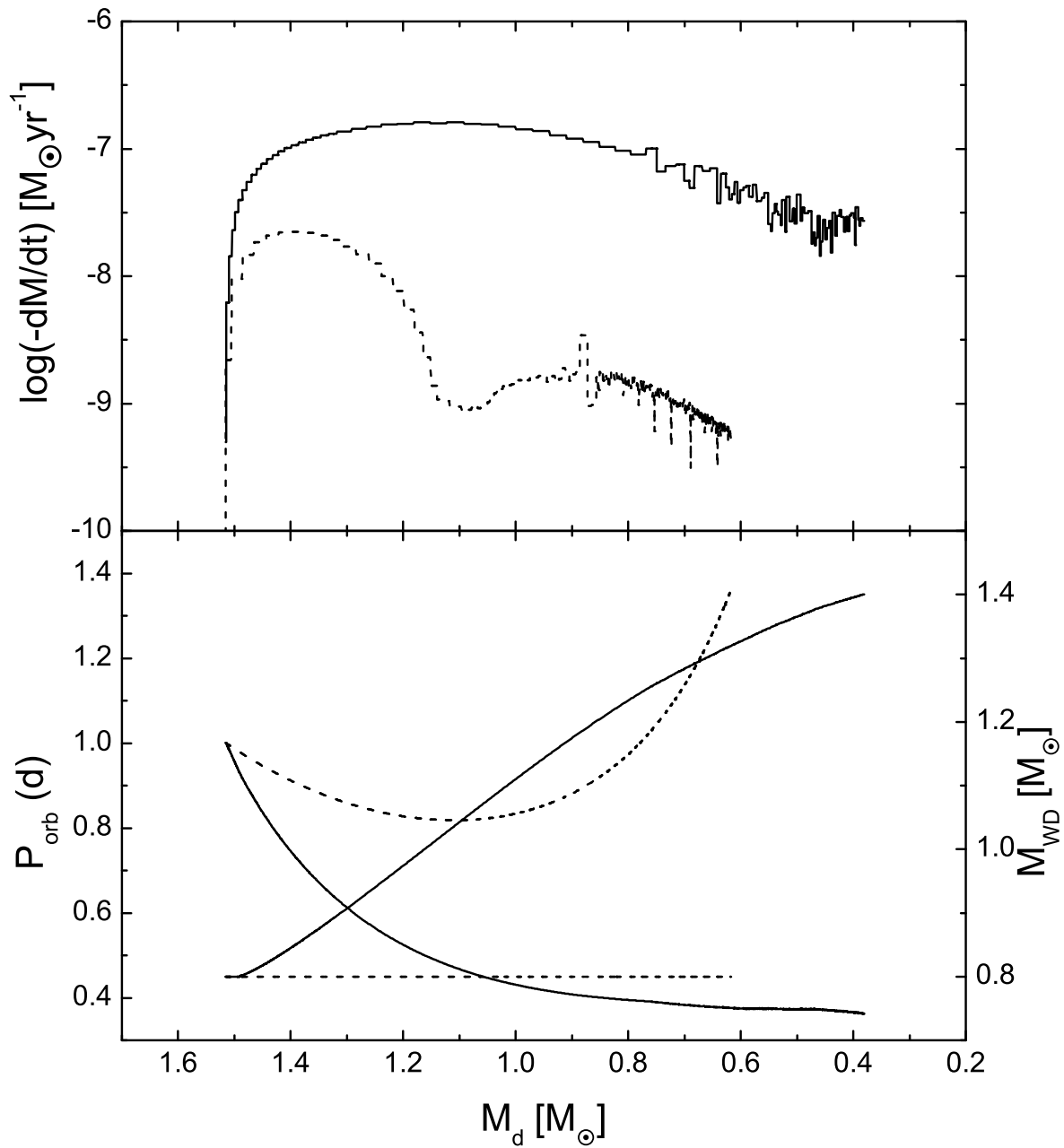} \caption{ Evolution of the mass
transfer rate $\dot{M_{\rm d}}$, orbital period $P_{\rm orb}$, and
WD mass $M_{\rm WD}$ (lower curves) for a WD binary with $M_{\rm
WD,i}=0.8M_{\odot}, M_{\rm d,i}=1.5M_{\odot}$ and $P_{\rm
orb,i}=1\rm d$. The solid and dashed curves denote the
evolutionary tracks with $\delta=0.01$ and 0, respectively.
\label{fig1}}
\end{figure}

\begin{figure}
\epsscale{.95} \plotone{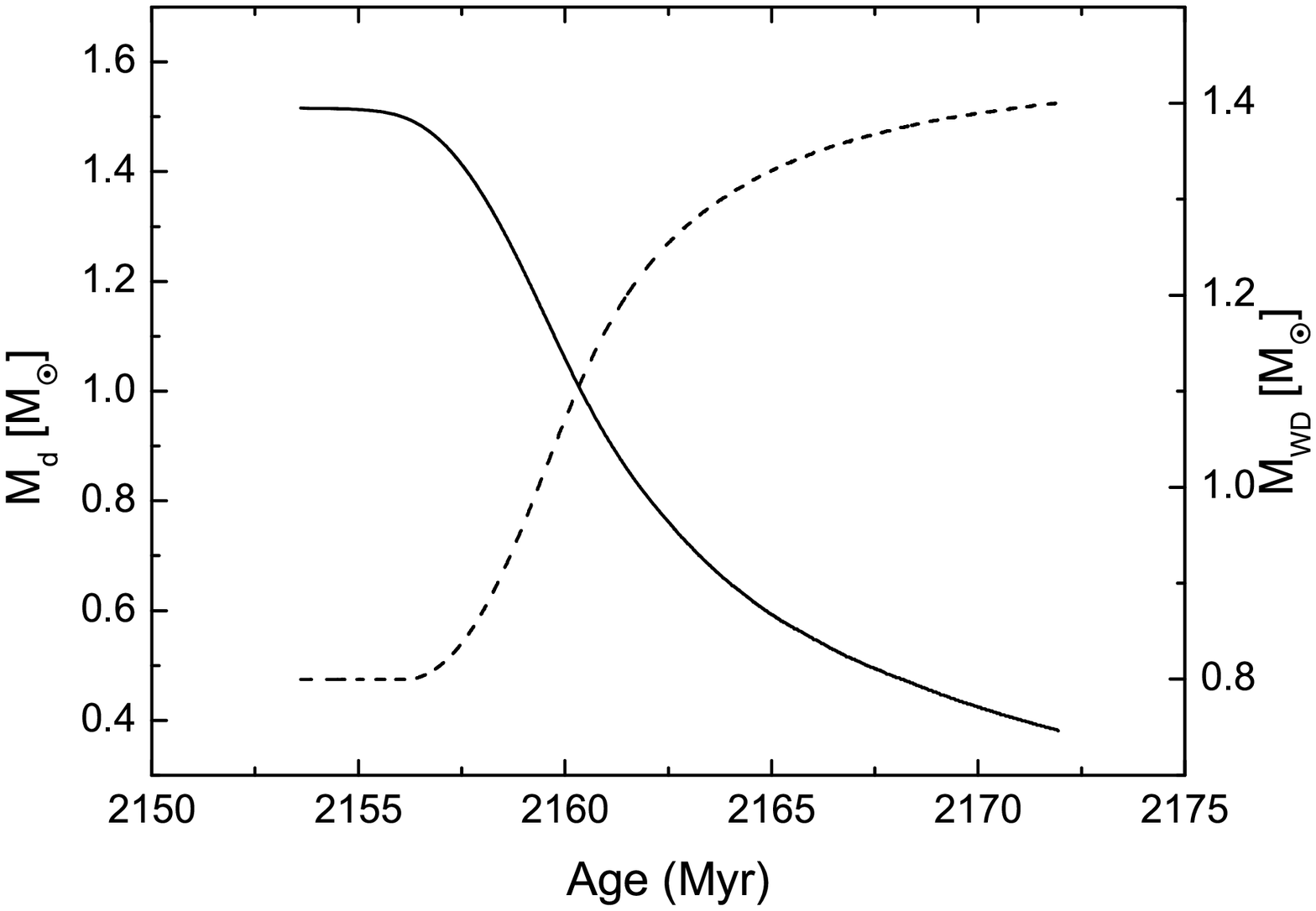} \caption{Evolution of the donor
mass (solid curve) and WD mass (dashed curve) when $M_{\rm
WD,i}=0.8M_{\odot}$, $M_{\rm d,i}=1.5M_{\odot}$, and $P_{\rm
orb,i}=1\rm d$. \label{fig2}}
\end{figure}

\section{Numerical Results}

We incorporate the prescriptions in last section into the stellar
evolution code, and calculate the evolution of WD binaries with
initial parameters of the WD mass $M_{\rm WD,i}$, donor mass $M_{\rm
d,i}$, and orbital period $P_{\rm orb,i}$. Once the WD mass $M_{\rm
WD}$ grows to $1.4M_{\odot}$, we stop the calculation and assume a
SNe Ia occurred. In our calculations, we set $\gamma=1.3,
\alpha_{\rm SS}=0.01$, and $\beta=0.03$ \citep{chen06b}.

Examples of the mass transfer sequences are shown in Figs.~1 and 2.
We plot the calculated evolutionary sequences for a system with
$M_{\rm WD,i}=0.8M_{\odot}$, $M_{\rm d,i}=1.5M_{\odot}$, and $P_{\rm
orb,i}=1$ d in Fig. 1. The solid and dashed curves correspond to the
cases of $\delta=0.01$ and 0, respectively. If no CB disk is assumed
to exist, the mass transfer rate is low enough
($\sim5\times10^{-10}-2\times10^{-8}M_{\odot}\rm yr^{-1}$) that the
accumulation ratio of the accreted matter $\alpha\sim 0$, and the WD
mass hardly increases. The orbital period first decreases to $\sim
0.8$ d, then increases to $\la 1.4$ d. When we include the effect of
the CB disk with $\delta=0.01$, the mass transfer rate maintains a
relatively high value in the range of
$\sim3\times10^{-8}-2\times10^{-7}M_{\odot}\rm yr^{-1}$. A large
fraction ($\sim55\%$) of the transferred material is accreted by the
WD, making the WD mass grow to $1.4M_{\odot}$ to trigger SNe Ia.
Note that the binary orbit secularly shrinks until the orbital
period $P_{\rm orb}$ reaches $\la 0.4$ d. This evolutionary sequence
may represent the formation history of systems like 1E 0035.4$-$7230
with narrow orbit, low-mass donor star and rapid mass transfer.
Figure 2 shows the evolution of the donor and WD masses with time.
It is clear that the donor star fills its Roche lobe when its age is
$\sim2$ Gyr for an initial donor star of $1.5M_{\odot}$, much longer
than the $\sim20$ Myr mass transfer time. This means that it is
possible for the SD/Ch model to have a few Gyr delay time.

We have calculated the evolutions of a large number of WD binaries
for a wide distribution of the initial input parameters. Figure 3
summarizes the final results of our binary evolution calculations
for the distributions of the progenitor systems of SNe Ia in the
$M_{\rm d,i}-P_{\rm orb,i}$ diagram. The bias and pane shading
denote the distribution area of WD binaries with a WD of initial
mass $M_{\rm WD,i}=1.2$ and $0.8M_{\odot}$, respectively. Beyond
these two areas, SNe Ia explosions cannot occur due to either low
mass accumulation rate or unstable mass transfer. Compared with
previous investigations, our CB disk model opens a possible
evolutionary channel to SNe Ia for WD binaries with relatively low
initial mass donor star of $\sim1-2\,M_{\odot}$ and hence long
delay times.

To investigate the properties of a companion star that has survived
the supernova explosion we have examined the consequences of heating
and mass stripping by the impact of the supernova shell in a SN Ia
on the secondaries. For the 24 ($M_{\rm WD,i}=0.8M_{\odot}$) and 71
($M_{\rm WD,i}=1.2M_{\odot}$) WD progenitor binaries shown in
Fig.~3, we calculate the mass loss fraction of the secondary star
following the semianalytical method of \citet{whee75}: the total
ejected mass fraction is given by $\eta_{\rm ej}=\eta_{\rm
st}+\lambda\eta_{\rm ev}$, where $\eta_{\rm st}$ and $\eta_{\rm ev}$
are the stripped and evaporated mass fraction respectively,
$\lambda$ is an adjusted parameter we assume to include the
uncertainties in estimating $\eta_{\rm ev}$. Here we take the mass
$M_{\rm SN}=1.4\,M_{\odot}$ and the velocity $v_{\rm SN}=8500$
kms$^{-1}$ for the SN ejecta \citep{mari00}, and $\lambda=0.5$. We
find that the subgiant secondary (of mass $0.3-2.6M_{\odot}$) loses
$\sim9\%-28 \%$ of its mass after the SN. Generally the greater mass
or binary separation, the smaller mass ejection fraction.
Figure 4 shows the distribution of the secondary stars in the HR
diagram just before and after the SN explosions. The latter could be
compared with and testified by future observations of the companions
of SNe Ia in young, nearby supernova remnants
\citep[e.g.][]{cana01}.
\begin{figure}
\epsscale{.95} \plotone{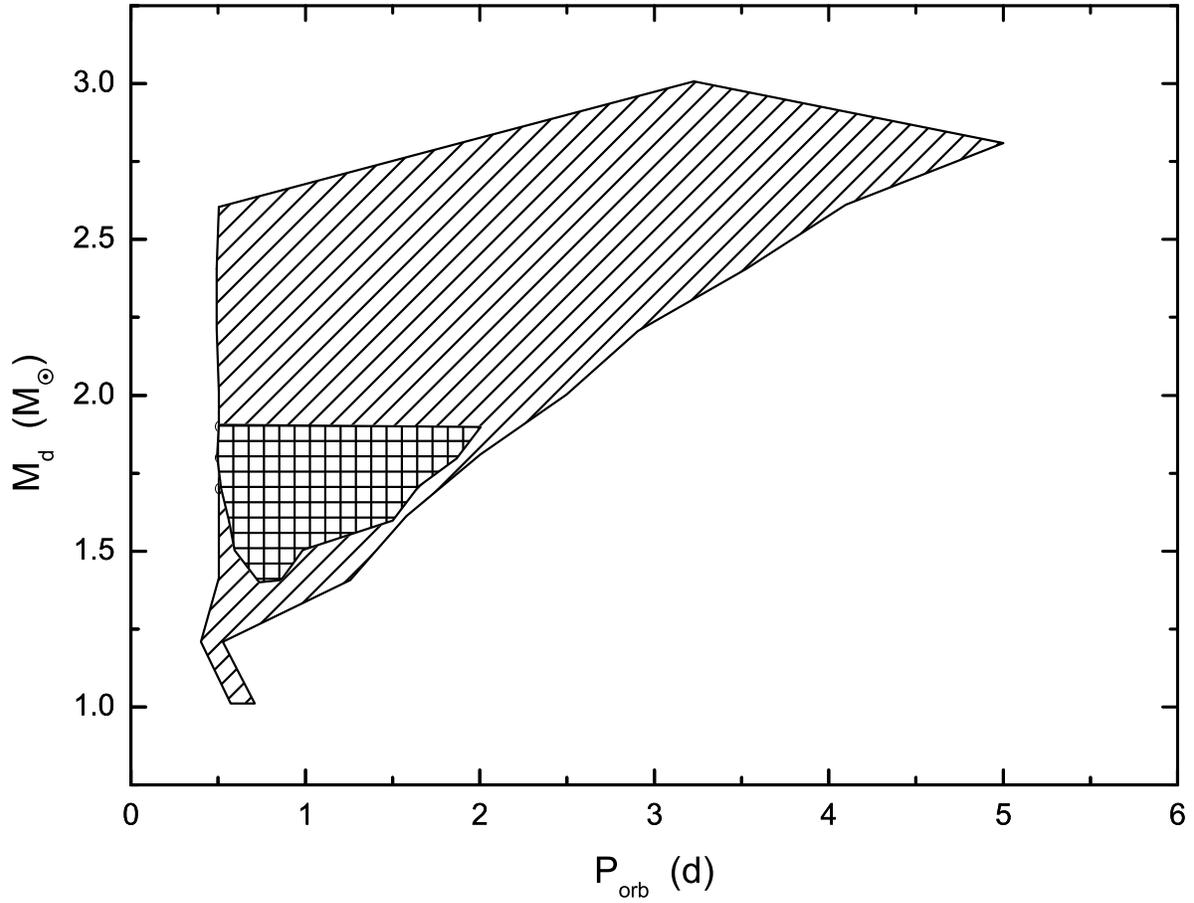} \caption{ Distribution of the
progenitor systems of SNe Ia in the $M_{\rm d,i}-P_{\rm orb,i}$
diagram. The bias and pane shading denote the distribution area of
WD binaries with a WD of initial mass $1.2M_{\odot}$ and
$0.8M_{\odot}$, respectively.\label{fig1}}
\end{figure}


\begin{figure}
\epsscale{.95} \plotone{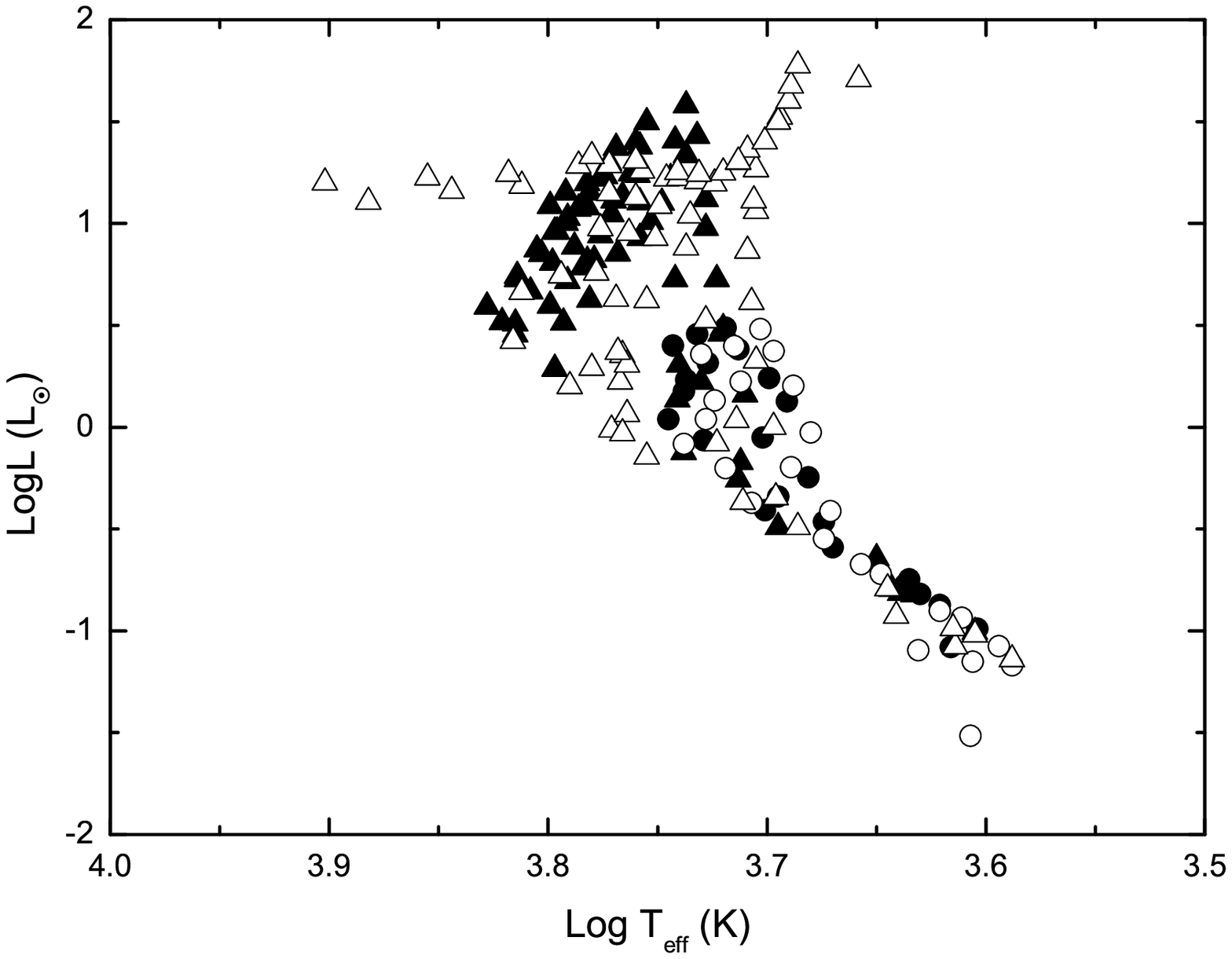} \caption{ Distribution of the
donor stars in the HR diagram when and after SNe Ia explosions
occur. The triangles and circles denote the donor stars
accompanied by an initial WD of $1.2M_{\odot}$ and $0.8M_{\odot}$,
while the solid and open signs represent the cases of when and
after SNe Ia explosions occur, respectively.\label{fig1}}
\end{figure}

\section{Discussion and Summary}
Recently \citet{mann06}, on the basis of observational arguments,
suggest that there is a bimodal delay time distribution, in which
about half of SNe Ia explode soon after their stellar birth, in a
time $\sim 10^8$ yr, while the remaining half have a much wider
distribution, well described by an exponential function with a
decay time of about 3 Gyr.  In the traditional SD/Ch model the
rapid mass transfer required limits the donor stars to be more
massive than $\sim 2M_{\odot}$ \citep{li97,lang00,han04}. The
innovation in this work is taking into account the effect of the
CB disk on the mass transfer process. Our calculations indicate
that the CB disk can significantly enhance the mass transfer
rates, especially for systems with low-mass ($<2\,M_{\odot}$)
donor stars, to trigger SNe Ia explosions. In particular, these
progenitor systems are consistent with the time delay of $\sim
1-3$ Gyr between the formation of the progenitors and the SNe Ia.
This scenario could also explain the existence of the peculiar SSS
containing rapidly accreting white dwarf and very low-mass donor
star like 1E 0035.4$-$7230 \citep[e.g.][]{taam01}.

Obviously there exist many uncertainties in the CB disk-driven
mass accretion onto WDs in this work. First, the existence of the
CB disk in supersoft X-ray binaries needs to be confirmed or
disproved by future infrared observations \citep{dubu04}, as in GG
Tau \citep{rodd96}. Second, the calculated mass transfer processes
depend sensitively on the adopted value of $\delta$, which is
poorly known, and likely to change with time. This makes it
difficult to estimate the contribution of the such binaries to SNe
Ia. Third, we do not include the influence of rotation of the
accreting WDs, which may cause the chandrasekhar mass to exceed
$1.4M_{\odot}$, i.e. the super-Ch mass model
\citep[e.g.][]{ueni03,yoon05,how06}. In conclusion, we want to
emphasize that before judging which kind of progenitor scenario
are more compatible with observations, we need to investigate
carefully various mass transfer processes in WD binaries.

\acknowledgments { This research was supported by Natural Science
Foundation of China under grant number 10573010, 10573009, and by
the Graduate Scientific Research Foundation of Jiangsu Province,
China.}

\end{document}